\documentclass[prd,preprint,showpacs,preprintnumbers,amsmath,amssymb]{revtex4}
\pdfoutput=1

\usepackage{mathrsfs}
\DeclareMathAlphabet{\mathpzc}{OT1}{pzc}{m}{it}

\usepackage[english]{babel}

\usepackage{amsmath}
\usepackage{amssymb}

\usepackage{mathrsfs}
\DeclareMathAlphabet{\mathpzc}{OT1}{pzc}{m}{it}

\usepackage{graphicx}

\usepackage{cancel} 
\usepackage{paralist}
\usepackage{booktabs}
\usepackage{hyperref}
\usepackage{xspace}

\usepackage{graphicx,color,verbatim,acronym,epsfig,subfigure}
\usepackage{slashed,skak}
\usepackage{latexsym,amsfonts,amssymb,amsmath,makeidx,mathrsfs,multirow,lineno,bm,bbm}

\begin{document}
\title{Gravitational waves from cosmic strings and first-order phase transition}

\author{Ruiyu Zhou,}
\email{zhoury@cqu.edu.cn}
\author{Ligong Bian \footnote{Corresponding author.}}
\email{lgbycl@cqu.edu.cn}

\affiliation{
	Department of Physics, Chongqing University, Chongqing 401331, China
}
\date{\today}

\begin{abstract}

Cosmic strings and first-order phase transition are two main sources for the stochastic gravitational wave background (SGWB).
In this work, we study the stochastic gravitational wave radiation from cosmic string which is formed after the first-order phase transition.
For the first-order phase transition occurs at temperature far beyond the electroweak scale, the gravitational wave signal cannot be reached by the future gravitational wave interferometers. The gravitational waves from cosmic strings that formed after the phase transition can be detected by future gravitational wave detectors in a wide range of frequency, and therefore its imprints can serve to search for firs-order phase transitions at high scales with the phase transition temperature: $ \mathcal{O}(10^8)$ GeV $ \leq T_n$ $\leq  \mathcal{O}(10^{11})$ GeV.  

\end{abstract}

\graphicspath{{figure/}}

\maketitle
\baselineskip=16pt

\pagenumbering{arabic}

\vspace{1.0cm}
\tableofcontents

\newpage

\section{Introduction}

The detection of gravitation waves (GWs) by LIGO/VIRGO~\cite{Abbott:2016blz,Abbott:2016nmj,Abbott:2017vtc,Abbott:2017gyy,Abbott:2017oio} raises people's growing interest on gravitational waves study, which provide an approach to probe the early universe and the new physics with high energy scale unaccessible by collider experiments.
The stochastic gravitational waves background (SGWB), as the preliminary targets at LIGO/VIRGO and LISA~\cite{Audley:2017drz,Auclair:2019wcv}, mainly sources from cosmological first-order phase transition with breaking power-law shape, cosmic string with scale-invariant shape, and inflation with near scale-invariant shape~\cite{Caprini:2018mtu}.
In this paper, we focus on the cosmic strings that are topological defects formed after the spontaneous symmetry breaking of U(1) symmetry in the early universe \cite{Hindmarsh:1994re,Vilenkin:2000jqa}. They are a generic prediction of beyond Standard Model theories, such as Grand Unified Theories \cite{King:2020hyd,Buchmuller:2019gfy,Buchmuller:2013lra}, or the seesaw mechanism where the $U(1)_{B-L}$ gets broken spontaneously \cite{Dror:2019syi,Okada:2020vvb,Blasi:2020wpy}.
In literatures, people mostly study the SGWB from cosmic strings disregarding the phase transition type that yields spontaneous symmetry breaking.  We propose to use the detection of a SGWB from local cosmic strings formed after spontaneous symmetry breaking driven by the first-order phase transition (FOPT).
We intent to explore the capability of the current and future GWs detectors to detect the SGWB from the FOPT and the produced cosmic strings. 

This paper is organized as follows: We first introduce the phase transition with the local $U(1)$ symmetry in Section.~\ref{sec:model}. The produced GWs from the FOPT and the cosmic strings are computed in Section.~\ref{sec:gw}.
Section.~\ref{sec:con} is devoted to the concluding remark.

\section{The Phase Transition model}
\label{sec:model}

The relevant Lagrangian is
\begin{equation}
\label{Lag1}
{\cal L} =  |D_\mu S|^2 - \frac{1}{4} F^\prime_{\mu\nu}F^{\prime\mu\nu} - V(S)\;,
\end{equation}
with $F^\prime_{\mu\nu}$ being the field strength tensors of
$U(1)^\prime$. The
covariant derivative is
\begin{align}
  D_\mu S &= \left(\partial_\mu + i g_D A_\mu^\prime\right)S \,,
\end{align}
where $g_D$ is the gauge coupling and $A_\mu^\prime$ is the gauge boson
of $U(1)'$.
The tree level scalar potential is given by
\begin{align}
  V_\text{tree}(S) &= - \mu_S^2 S^\dag S
  + \frac{\lambda_S}{2} (S^\dag S)^2\;,
  \label{eq:Vtree}
\end{align}
where the $\mu_S^2=\lambda_S v_s^2/2$ can be obtained with the minimal condition of the potential,
\begin{align}
~~\frac{dV_\text{tree}(S)}{ds}|_{s=v_s}=0\;.
\end{align}

With the standard methodology, the phase transition can be studied with the thermal one-loop effective potential~\cite{Quiros:1999jp},
\begin{align}
V_{eff}(s,T)=V_{\rm 0} (s)+ V_{CW}(s)+V^{\rm c.t}_{1}(s)+ V_{1}^{T}(s,T)\, .
\end{align}
The $V_{\rm 0} (s)$ is the tree-level potential for the classical field,
\begin{equation}
V_{0} (s) =- \frac{\mu_s^2}{2}s^2 + \frac{\lambda_s}{8} s^4.
\end{equation}
The Coleman-Weinberg contribution is given by~\cite{Coleman:1973jx}	
\begin{align}
	V_{CW}(s)= \sum_{i} \frac{g_{i}(-1)^{F}}{64\pi^2}  m_{i}^{4}(s)\left(\mathrm{Ln}\left[ \frac{m_{i}^{2}(s)}{\mu^2} \right] - C_i\right)\,,
	\label{eq:oneloop}
\end{align}
Where,  $F=0 \; (1)$ for bosons (fermions), $\mu$ is the $\overline{\text{MS}}$ renormalization scale, $g_{i}= \{1,3,1 \}$ for the $s, \; A'$, $G^D$ in $U(1)$ model, and $C_i=5/6$ for gauge bosons and $C_i=3/2$ for scalar fields and fermions.
The field-dependent masses of the scalar, Goldstone, and of the gauge boson are
\begin{align}
  m^2_{S}(s)   = -\mu_S^2+\frac{3}{2}\lambda_S s^2 \,,~~
  m^2_{G_D}(s) = -\mu_S^2+\frac{1}{2}\lambda_S s^2 \,,~~
  m^2_{A'}(s)       = g_D^2 s^2 \,.
\end{align}

The counter terms to the potential in Eq.~\eqref{eq:oneloop} are
\begin{equation}
V^{\rm c.t}_{1} (s) = -\frac{\delta\mu_2^2}{2} s^{2} + \frac{\delta\lambda_2}{8}s^{4}.
\label{eq:CT}
\end{equation}
To prevent shifts of the masses and VEVs of the scalars from their tree level values, we impose
\begin{eqnarray}
\partial_{s} (V_{CW}(s) + V^{\rm c.t}_{1}(s)) \bigg|_{s=v_s}&=& 0\,,\nonumber\\
\partial_{s} \partial_{s} (V_{CW}(s) + V^{\rm c.t}_{1}(s)) \bigg|_{s=v_s}&=& 0\,.
\end{eqnarray}

The finite temperature effective potential at one-loop is given by
\begin{eqnarray}
V_{1}^{T}(s, T)= \frac{T^4}{2\pi^2}\, \sum_i g_i J_{B,F}\left( \frac{ M_i^2(s)+\Pi_i(T)}{T^2}\right),
\end{eqnarray}
 the Debye masses are calculated as
\begin{align}
  \Pi_{S(G)}(T)  = \left(\frac{\lambda_S}{6} + \frac{g_D^2}{4}\right) T^2 \,, ~~~
  \Pi_{A'}(T) = \frac{g_D^2}{3} T^2 \,.
\end{align}
The functions $J_{B,F}(y)$ are
\begin{eqnarray}
 J_{B,F}(y) = \pm \int_0^\infty\, dx\, x^2\, \ln\left[1\mp {\rm exp}\left(-\sqrt{x^2+y}\right)\right]\;,
\end{eqnarray}
Where, the upper (lower) sign corresponds to bosonic (fermionic) contributions. The above integral $J_{B,F}$ can be expressed as a sum of them second kind modified Bessel functions $K_{2} (x)$~\cite{Bernon:2017jgv},
\begin{eqnarray}
J_{B,F}(y) = \lim_{N \to +\infty} \mp \sum_{l=1}^{N} {(\pm1)^{l}  y \over l^{2}} K_{2} (\sqrt{y} l)\;.
\end{eqnarray}

With the thermal effective potential obtained above, one can get
the solution of the bounce configuration
(the bounce configuration of the field connects the $U(1)^\prime$ broken vacuum (true vacuum,) and the $U(1)^\prime$ preserving vacuum (false vacuum)) of the nucleated bubble, which is obtained by extremizing,
\begin{eqnarray}
S_3(T)=\int 4\pi r^2d r\bigg[\frac{1}{2}\big(\frac{d \phi_b}{dr}\big)^2+V_{eff}(\phi_b,T)\bigg]\;,
\end{eqnarray}
after solving the equation of motion for the field $\phi_b$,
which satisfies the boundary conditions
\begin{eqnarray}
\lim_{r\rightarrow \infty}\phi_b =0\;, \quad \quad \frac{d\phi_b}{d r}|_{r=0}=0\;.
\end{eqnarray}
The $\phi_b$ is the $s$ field considered in this work.
The nucleation temperature ($T_n$) is obtained when the thermal tunneling probability for bubble nucleation per horizon volume and per horizon time is of order unity~\cite{Affleck:1980ac,Linde:1981zj,Linde:1980tt}:
\begin{eqnarray}\label{eq:bn}
\Gamma\approx A(T)e^{-S_3/T}\sim 1\;,
\end{eqnarray}
where the phase transition completes.

\section{Gravitational waves}
\label{sec:gw}

We calculate the GWs from FOPT, and the local cosmic string produced after the spontaneous symmetry breaking at high scales.

\subsection{ GWs from FOPT}

One crucial parameter for the calculation of the gravitational wave is the energy budget of the FOPT normalized by the radiative energy, which is defined as~\cite{Caprini:2015zlo}
\begin{eqnarray}
\alpha=\frac{\Delta\rho}{\rho_R}\;.
\end{eqnarray}
Here, the $\Delta \rho$ is the released latent heat from the phase transition to the energy density of the plasma background. Another crucial parameter $\beta$ characterizing the inverse time duration of the phase transition, which is given as
\begin{eqnarray}
\frac{\beta}{H_n}=T\frac{d (S_3(T)/T)}{d T}|_{T=T_n}\; .
\end{eqnarray}

The gravitational waves from the FOPT mainly include two sources: sound waves, and MHD turbulence, with the total energy being given by~\cite{Caprini:2015zlo}
\begin{equation}
\Omega_{\rm GW} h^2 (f) \approx \Omega h^2_{\rm sw} (f) +\Omega h^2_{\rm turb}(f).
\end{equation}
The detonation bubble is adopted and the bubble wall velocity $v_b$ is a function of $\alpha$~\cite{Steinhardt:1981ct}\footnote{We note that to compatible with EWBG, the wall velocity here can be obtained as a function of $\alpha$.~\cite{Zhou:2020idp,Zhou:2019uzq,Zhou:2020xqi,Alves:2018oct,Alves:2018jsw,Alves:2019igs} after taking into account Hydrodynamics. },
\begin{equation}\label{eq:bubblespeed}
v_b=\frac{1/ \sqrt{3}+\sqrt{\alpha ^2+2 \alpha /3}}{1+\alpha }\;.
\end{equation}

Sound waves created in the plasma constitute the leading source of GWs,
its energy density is given by
\begin{equation}
\Omega h^2_{\rm sw}(f)=2.65 \times 10^{-6}(H_*\tau_{sw})\left(\frac{\beta}{H}\right)^{-1} v_b
\left(\frac{\kappa_\nu \alpha }{1+\alpha }\right)^2
\left(\frac{g_*}{100}\right)^{-\frac{1}{3}}
\left(\frac{f}{f_{\rm sw}}\right)^3 \left(\frac{7}{4+3 \left(f/f_{\rm sw}\right)^2}\right)^{7/2},
\end{equation}
where the $\tau_{sw}=min\left[\frac{1}{H_*},\frac{R_*}{\bar{U}_f}\right]$, $H_*R_*=v_b(8\pi)^{1/3}(\beta/H)^{-1}$ is to consider the duration of the phase transition~\cite{Ellis:2020awk}. The root-mean-square (RMS) fluid velocity can be approximated as \cite{Hindmarsh:2017gnf, Caprini:2019egz, Ellis:2019oqb}
\begin{equation}
\bar{U}_f^2\approx\frac{3}{4}\frac{\kappa_\nu\alpha}{1+\alpha}\;.
\end{equation}
The term $H_*\tau_{\rm sw}$ accounts for the GW amplitude for sound wave suppressed by a factor of $H_*R_*/\overline{U}_f$, if the sound wave source can not last more than a Hubble time. The fraction of the latent heat transferred into the kinetic energy of plasma is described by the $\kappa_\nu$ factor, which can be obtained through considering the the hydrodynamic analysis~\cite{Espinosa:2010hh}.
The peak frequency locates at~\cite{Hindmarsh:2013xza,Hindmarsh:2015qta,Hindmarsh:2017gnf}
\begin{equation}
f_{\rm sw}=1.9 \times 10^{-5} \frac{\beta}{H} \frac{1}{v_b} \frac{T_*}{100}\left({\frac{g_*}{100}}\right)^{\frac{1}{6}} {\rm Hz }\;.
\end{equation}
The MHD turbulence in the plasma is the sub-leading source of GW signals,
with the energy density being given by
\begin{equation}
\Omega h^2_{\rm turb}(f)=3.35 \times 10^{-4}\left(\frac{\beta}{H}\right)^{-1}
\left(\frac{\epsilon \kappa_\nu \alpha }{1+\alpha }\right)^{\frac{3}{2}}
\left(\frac{g_*}{100}\right)^{-\frac{1}{3}}
v_b
\frac{\left(f/f_{\rm turb}\right)^3\left(1+f/f_{\rm turb}\right)^{-\frac{11}{3}}}{\left[1+8\pi f a_0/(a_* H_*)\right]}\;,
\end{equation}
with peak frequency locating at \cite{Caprini:2009yp}
\begin{equation}f_{\rm turb}=2.7  \times 10^{-5}
\frac{\beta}{H} \frac{1}{v_b} \frac{T_*}{100}\left({\frac{g_*}{100}}\right)^{\frac{1}{6}} {\rm Hz }\;.
\end{equation}
The efficiency factor $\epsilon \approx 0.1$, and the present Hubble parameter is obtained as
\begin{equation}
	H_{\ast} = \bigl( 1.65 \times 10^{-5} Hz \bigr) \left( \frac{T_{*}}{100 \rm{GeV}} \right) \left( \frac{g_{\ast}}{100} \right)^{1/6}\;.
\end{equation}
For this study, we consider $T_\star\approx T_n$.

\subsection{ GWs from cosmic strings}

In this paper, we consider Nambu-Goto cosmic strings characterized solely by the string tension $\mu$, with string tension $\mu\approx 2\pi v_s^2\;n$ with $n$ being winding number~\cite{Vilenkin:2000jqa}. 
Following the
Kibble mechanism, which can be roughly estimated to be~\cite{Gouttenoire:2019kij,Gouttenoire:2019rtn}
 \begin{eqnarray}
 \mu\approx \frac{10^{-15}}{\rm G}\left(\frac{T_p}{10^{11}~ {\rm GeV}}\right)^2\;,\label{eq:ten}
 \end{eqnarray}
 where G is Newton's constant. We adopte $T_p\approx T_n$ in this study. After formation, the string loops loss energy dominantly through emission of gravitational waves.
We calculate the the relic GW energy density spectrum from cosmic string networks following Ref.~\cite{Cui:2018rwi},
\begin{eqnarray}
\label{eq:GWdensity1}
\Omega_{\rm GW}(f) =\sum_k \Omega_{\rm GW}^{(k)}(f)\; ,
\end{eqnarray}
with k-mode being
\begin{eqnarray}
\label{eq:GWdensity2}
\Omega_{\rm GW}^{(k)}(f) =
\frac{1}{\rho_c}
\frac{2k}{f}
\frac{\mathcal{F}_{\alpha}\,\Gamma^{(k)}G\mu^2}
{\alpha\left( \alpha+\Gamma G\mu\right)}
\int_{t_F}^{t_0}\!d\tilde{t}\;
\frac{C_{eff}(t_i^{(k)})}{t_i^{(k)\,4}}
\bigg[\frac{a(\tilde{t})}{a(t_0)}\bigg]^5
\bigg[\frac{a(t^{(k)}_i)}{a(\tilde{t})}\bigg]^3
\,\Theta(t_i^{(k)} - t_F)~~~~~
\end{eqnarray}
Here, $\rho_c = 3H_0^2/8\pi G$ is the critical density,
the factor $\mathcal{F}_{\alpha}$ characterizes the fraction of the energy released by long strings and we take $\mathcal{F}_{\alpha}=0.1$,
 $\alpha = 0.1$ is adopted to consider the length of the string loops considering a monochromatic loop distribution.
 The loop production efficiency $C_{eff}$ is obtained after solving Velocity-dependent One-Scale equations (VOS),  with $C_{eff}=5.4(0.39)$ in radiation (matter) dominate universe~\cite{Gouttenoire:2019kij}\footnote{Here, we note that the scaling regime is reached after 3 or 4 orders of magnitude of change in the energy scale of the universe, where we have stable $C_{eff}$, see Fig.3 of Ref.~\cite{Gouttenoire:2019kij}.}.
The gravitational loop-emission efficiency $\Gamma\approx50$~\cite{Blanco-Pillado:2017oxo} with its Fourier modes for cusps~\cite{Olmez:2010bi}
(emission rate per mode) being given by\cite{Blanco-Pillado:2013qja,Blanco-Pillado:2017oxo}:
\begin{equation}\label{eq:Gammak}
\Gamma^{(k)} = \frac{\Gamma k^{-\frac{4}{3}}}{\sum_{m=1}^{\infty} m^{-\frac{4}{3}} } \;,
\end{equation}
here, $\sum_{m=1}^{\infty} m^{-\frac{4}{3}} \simeq 3.60$ and
$\sum_k \Gamma^{(k)}$.
The formation time of loops of the k mode is a function of the GW emission time $\tilde{t}$, casts the form of
\begin{equation}\label{eq:ti}
t_i^{(k)}(\tilde{t},f) = \frac{1}{\alpha+\Gamma G\mu}\left[
\frac{2 k}{f}\frac{a(\tilde{t})}{a(t_0)} + \Gamma G\mu\;\tilde{t}\;
\right].
\end{equation}
  The cosmic string network reaches scaling after formation at time $t_F$, which connect with the phase transition through~\cite{Gouttenoire:2019kij}:
  \begin{equation} 
  \sqrt{\rho_{tot}(t_F)}\equiv \mu\;.\label{eq:tf}
  \end{equation}
For the case where the small-scale structure of loops is dominated by cusps, the high mode in Eq.~\eqref{eq:GWdensity1} can be evaluated as
$\Omega_{\rm GW}^{(k)}(f)
= \frac{\Gamma^{(k)}}{\Gamma^{(1)}}\,\Omega_{\rm GW}^{(1)}(f/k)
=k^{-4/3}\,\Omega_{\rm GW}^{(1)}(f/k)$\;.
The low and high frequencies of the spectrum of the GWs from cosmic strings are dominated by emissions in matter dominate and radiation dominate universe. The spectrum at high frequencies is flat in a wide range, which is therefore expected to be probed utilizing the complementary searches at LIGO~\cite{Abbott:2016blz,Aasi:2014mqd,Thrane:2013oya,LIGOScientific:2019vic}, SKA~\cite{Janssen:2014dka}, EPTA~\cite{Desvignes:2016yex}, PPTA~\cite{Hobbs:2013aka}, IPTA~\cite{Verbiest:2016vem}, TianQin~\cite{Luo:2015ght}, Taiji~\cite{Gong:2014mca}, LISA~\cite{Audley:2017drz}, Einstein Telescope \cite{Hild:2010id, Punturo:2010zz}, Cosmic Explorer~\cite{Evans:2016mbw}, BBO~\cite{Corbin:2005ny} and DECIGO~\cite{Yagi:2011wg}.

\subsection{Numerical results}

\begin{table}[htp]
\begin{center}
\begin{tabular}{c c c c c c c c c }
\hline
~Benchmark points~&~$\lambda$~&~~~$g_D$~~~ &~$v_s$~(GeV)~ & ~$v_n$~(GeV)~ & ~ $T_n$~(GeV)~ & ~ $\alpha$~ & ~ $\beta/H_n$  \\
\hline
$BM_1$  & $0.096$ & $1.459$ & $1.056\times 10^{9}$ & $1.051\times 10^{9}$ & $2.545\times 10^{8}$ & $0.031$ & $6.467\times 10^{4}$ \\

$BM_2$  & $0.822$ &$2.164$&$1.921\times 10^{10}$&$1.774\times 10^{10}$&$1.332\times 10^{10}$& $0.015$ & $3.005\times 10^{6}$ \\

\hline
\end{tabular}
\caption{The six benchmark points in Fig.~\ref{GW_2}.}
\label{tabp2}
\end{center}
\end{table}

\begin{figure}[!htp]
\begin{center}
\includegraphics[width=0.6\textwidth]{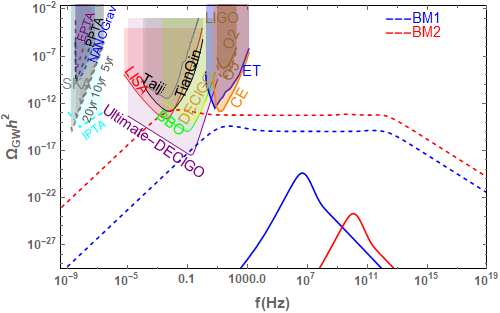}
\caption{The gravitational waves from cosmic strings are shown in dashed lines, and gravitational waves from first-order phase transitions considering duration of the phase transition are shown in solid lines.}
\label{GW_2}
\end{center}
\end{figure}

The CMB measurements bounds
 the dimension less parameter to be $G\mu \leq 1.1 \times 10^{-7}$~\cite{Charnock:2016nzm}.
 The current strongest constraints come from pulsar timing array EPTA and NANOGrav: $G\mu\leq 8\times10^ {-10}$ \cite{Lentati:2015qwp}, $G\mu\leq 5.3\times10^ {-11}$~\cite{Arzoumanian:2018saf}, the $G\mu< 10^{-11}$ corresponds to $T_n \leq 10^{15}$ GeV for FOPT scenarios. 
 With the Eq.~(\ref{eq:ten},\ref{eq:tf}), we found that one cannot have formation of string loops for the $T_n\geq \mathcal{O}(10^{11} ) {\rm GeV}$. 
 With the Fig.~\ref{GW_2}, we present the GW spectrum contributed from cosmic string networks and FOPT for benchmarks shown in Table.~\ref{tabp2}. 
The GWs from the
FOPT at high frequency, corresponding to symmetry breaking scale such as $v_s\geq 10^9$ GeV, is found to beyond capacities of any GWs detectors. The magnitude of the GWs from cosmic strings would be higher for higher phase transition temperatures. The cuttoff effects shown in the spectrum of the GWs from cosmic strings is beyond ranges of any detectors. It correspond to the formation of string loops, and the phase transition temperature for FOPT studied in this paper. Considering the detectability of future interferometers, the lowest value $G\mu\geq\mathcal{O}(10^{-18})$ corresponds to phase transition temperature $T_n\geq \mathcal{O}(10^8)$ GeV. 
We further note that LISA sensitivity to the GWs from cosmic strings shows that~\cite{Auclair:2019wcv}: LISA is capable to probe cosmic string with tensions $G\mu\geq \mathcal{O}(10^{-17})$, considering Nambu-Goto strings with the average loop size at formation being $\alpha\approx 0.1$. Therefore, we expect the 
GWs from cosmic strings from FOPT with $T_n\geq\mathcal{O}(10^{10}) {\rm GeV}$ studied in this paper can be probed by LISA, as shown by the red curved in Fig.~\ref{GW_2}.

\section{Conclusions and discussions}
\label{sec:con}

In this paper, we study the scenario where the stochastic gravitational wave background coming from both FOPT and the
cosmic string networks produced after the abelian $U(1)^\prime$ symmetry is broken. For the scenario where the SGWB from the cosmic strings can be probed by the GW detectors, the peak frequency of the SGWB contribution from the FOPT is beyond the sensitivity regions of these detectors. Considering capabilities of all the future GW detectors, we found the SGWB from the cosmic strings
produced after the FOPT with $ \mathcal{O}(10^8)$ GeV $ \leq T_n$ $\leq  \mathcal{O}(10^{11})$ GeV can be detected, LISA is able to probe the GWs from the cosmic strings with the first-order phase transition temperature lives in a quite narrow range: $ \mathcal{O}(10^8)$ GeV $ \leq T_n$ $\leq  \mathcal{O}(10^{11})$ GeV.

We didn't consider the supercooling phase transition, where one may have slightly stronger GWs from the FOPT~\cite{Ellis:2019oqb,Ellis:2018mja,Wang:2020jrd}. While since the string tensions is determined by the phase transition temperature, our study on the SGWB from cosmic strings still apply.
The loop-production efficiency $C_{eff}$, that is crucial for the magnitude of the gravitational waves spectrum, obtained for Nambu-Goto simulations~\cite{Martins:2000cs} is about four times larger than the Abelian-Higgs simulations~\cite{Moore:2001px}, which may overestimate the SGWB.
To set if the Nambu-Goto approximation applied for the Abelian-Higgs theory requires quantum field string lattice simulations for the theory, which is still in debate, see Ref.~\cite{Vincent:1997cx,Hindmarsh:2008dw,Hindmarsh:2017qff} and Ref.~\cite{Moore:1998gp,Olum:1999sg,Moore:2001px}. For the details on the mass radiation
 and thermal frictions from the cosmic strings loops we refer to Ref.~\cite{Gouttenoire:2019kij}. The interplay between the phase transitions and the cosmic strings request future simulation, which may set the scaling regime and the phase transition temperature, and therefore set the ability of future interferometers sensitivities to cosmic strings from FOPT.

\section{Acknowledgements}

This work is supported by the National Natural Science Foundation of China under grant No.11605016 and No.11947406, and the Fundamental Research Funds for the Central Universities of China (No. 2019CDXYWL0029).
We are grateful to Alexander Vilenkin, Yanou Cui, Marek Lewicki, David E. Morrissey, James D. Wells, Daniel G. Figueroa, Ruth Durrer, David Weir, Jeff A. Dror, Yann Gouttenoire, Luca Visinelli, Nicklas Ramberg, Eric Madge, Huai-Ke Guo, Yue Zhao, and Ye-Ling Zhou for helpful communications and discussions.

\newpage

\bibliographystyle{unsrt}

\end{document}